\documentstyle[12pt]{article}

\begin{document}
\begin{titlepage}
\title{FINITE VOLUME OF BARYONS AND THE MASS LIMIT FOR NEUTRON STARS}
\author{Qi-Ren Zhang\thanks{On leave from the Department of Technical
Physics, Peking University, Beijing 100871, China},
Bo-Qiang Ma\thanks{Fellow of Alexander Von Humboldt foundation},
and Walter Greiner\\
Institute for Theoretical Physics\\
Johann Wolfgang Goethe University\\
D-6000 Frankfurt/Main, Germany}
\date{}
\maketitle
\begin{abstract}
We confirmed the following idea by numerical calculation: The extended
structure of baryons may make the nuclear equation of state stiffer at
higher density while keeping the compression modulus for normal nuclear
matter around its empirical value $K=240MeV$. The model built in this way
fits all empirical data for normal nuclear matter and gives the mass
limit of neutron stars compatible with the observation.

\noindent
To be published in J.Phys.G
\end{abstract}
\end{titlepage}

\begin{center}
I. Introduction
\end{center}

It was found by Glendenning\cite{g1} that
from the local relativistic mean field theory
a nuclear equation of state with compression modulus $K$ around
$200MeV$ gives too low a mass limit of neutron stars, lower than the mass
$1.85M_\odot$ of the observed neutron star 4U0900-40. The
introduction
of a derivative coupling
between baryons and the scalar meson field\cite{g2} improves the result,
but it destroys the renormalizability which is the main reason to keep
the field theory in its local form. On the other hand, it is widely
accepted that the hadrons, especially the baryons, are extended objects.
It means if consider them by field theory we have to cut the volumes
occupied by their extended structure out from their configuration space.
The effect of these finite volumes is similar to
that of hardcores for baryons.
The idea of hardcores for baryons comes in here naturally. This idea has
been successfully used in nuclear physics for decades, and is still
successfully used in recent works\cite{b,l,z,g}. In the theory of nuclear
matter the effect of hardcore is usually considered by the Van der Waals
approximation\cite{b,l,z,g}. The problem of thermodynamical consistency
has been carefully examined\cite{g}. The influence of hardcores becomes
more important for nuclear matter at higher densities. If we fix model
parameters by fitting normal nuclear matter data, the equation of state
at high densities for hadron matter made of hardcore baryons will be
stiffer than that made of point baryons. This idea makes us believe that
after considering the finite volume effect we may fit both the normal
nuclear data including the compression modulus and the data at higher
nuclear densities, including neutron star data and the data of heavy ion
collisions. The motivation of this work is to confirm the idea by
numerical calculation for the mass limit of neutron stars. The result is
encouraging. From the theory of the nuclear force we know that a momentum
dependent potential is equivalent to a hardcore in some sense. This point
helps us to understand why a derivative coupling between nucleons
and the scalar meson field improves the result.

In section II, we formulate the problem and determine the model parameters
by fitting normal nuclear matter data. In section III, we derive the
equation of state for neutron star matter.
In section IV we solve the Oppenheimer-Volkov  equation numerically
to find the mass limit for the neutron stars.
The calculated mass limit is indeed larger than the mass of
observed heaviest neutron star. Section V is a discussion.

\begin{center}
II. Normal nuclear matter and the model parameters
\vskip 0.1in

A. Point nucleons interacting with meson fields
\end{center}

Consider a system composed of the local nucleon field $\Psi$ ,
the scalar meson
field $\phi$, the $\omega$ meson field $V_\mu$, and the $\rho$ meson
field $V_\mu^{(k)}$. Its Lagrangian density may be written in the form
\begin{eqnarray}
{\cal L} &=& -\bar{\Psi}[\gamma_\mu(\partial_\mu-{\rm i}g_\omega V_\mu
-{\rm i}g_\rho \sum_{k=1}^3 \tau^{(k) }V_\mu^{(k)})+m-g\phi]\Psi
\nonumber \\
         & &-\frac{1}{4}(F_{\mu\nu}F_{\mu\nu}+\sum_{k=1}^3F_{\mu\nu}^{(k)}
F_{\mu\nu}^{(k)})-\frac{1}{2}
(\partial_\mu\phi\partial_\mu\phi \nonumber  \\
         & & +m_\omega^2 V_\mu V_\mu + m_\rho^2
\sum_{k=1}^3 V_\mu^{(k)}V_\mu^{(k)})-U(\phi),
\label{1}
\end{eqnarray}
in which $\partial_\mu=\frac{\partial}{\partial x_\mu}, \tau^{(k)}$ is
the Pauli matrix for isotopic spin, $m, m_\omega,$ and $m_\rho$ are masses
for nucleon, $\omega$ meson and $\rho$ meson respectively, $g_\omega,
g_\rho$ and $g$ are coupling constants of nucleon with $\omega, \rho,$
and scalar mesons, $U(\phi)$ is the potential energy density for scalar
meson field, usually is a polynomial of degree 4 in $\phi$,
\begin{eqnarray}
F_{\mu\nu}
= \partial_\mu V_\nu -\partial_\nu V_\mu, \;\;\;\; F_{\mu\nu}^{(k)}=
 \partial_\mu V_\nu^{(k)} -\partial_\nu V_\mu^{(k)}. \label{2}
\end{eqnarray}
Einstein convention for summation over repeated Greek indices
is assumed here.

Extremizing the action
\begin{eqnarray}
J=\int {\cal L}\, d^4x                                  \label{3}
\end{eqnarray}
with respect to $V_4$ and $V_4^{(k)}$, we find
\begin{eqnarray}
\bigtriangledown \cdot {\bf E} =-m_\omega^2V_0 +g_\omega n,\;\;\;
\bigtriangledown \cdot {\bf E}^{(k)} =-m_\rho^2 V_0^{(k)} +g_\rho n^{(k)}
,\label{4}
\end{eqnarray}
in which
\begin{eqnarray}
{\bf E}
= -\bigtriangledown V_0 - \frac{\partial {\bf V}}
{\partial x^0}\;\;\;\;\mbox{and}\;\;\;\;
{\bf E}^{(k)}
= -\bigtriangledown V_0^{(k)} - \frac{\partial {\bf V}^{(k)}}
{\partial x^0} \label{5}
\end{eqnarray}
are the $\omega-$ and $\rho-$ electric field strength respectively, and
\begin{eqnarray}
n=\Psi^\dagger \Psi  \;\;\;\;\;\;\mbox{and} \;\;\;\;\;\;
n^{(k)}=\Psi^\dagger \tau^{(k)} \Psi  \label{6}
\end{eqnarray}
are the density and isotopic spin density for nucleons respectively.
$V_0$ and $V_0^{(k)}$ may be solved from(\ref{4}) and (\ref{5}) as
functions of ${\bf V},{\bf V}^{(k)}$ and $\Psi$, therefore are not
independent dynamical variables. They may be eliminated from the formalism.
By the standard canonical method we find the Hamiltonian
\begin{eqnarray}
H=\int {\cal H}\, d^3 x, \label{7}
\end{eqnarray}
with the Hamiltonian density
\begin{eqnarray}
{\cal H}\!&\!=\!&\! \Psi^\dagger [\vec \alpha
\cdot (-{\rm i}\bigtriangledown)
+m-g\phi ]\Psi +\frac{1}{2}[(\frac{\partial \phi}{\partial x^0})^2
+(\bigtriangledown \phi)^2 ]+U(\phi)   \nonumber \\
  &\!+\!&\! \frac{1}{2}[E^2 +B^2 +m_\omega^2 V^2 +\frac{1}{m_\omega^2}(
\bigtriangledown
\cdot {\bf E}-g_\omega n)^2]-g_\omega {\bf V}\cdot {\bf j}  \nonumber \\
  &\!+\!&\! \sum_{k=1}^3 \{ \frac{1}{2}[(E^{(k)})^2 + (B^{(k)})^2+m_\rho^2
(V^{(k)})^2+\frac{1}{m_\rho^2}(\bigtriangledown\cdot{\bf E}^{(k)}
-g_\rho n^{(k)})^2]\nonumber
\\ &\!-\!&\!g_\rho{\bf V}^{(k)}\cdot{\bf j}^{(k)
\end{eqnarray}
\begin{eqnarray}
{\bf B}=\bigtriangledown \times {\bf V}
\;\;\;\;\mbox{and}\;\;\;\;
{\bf B}^{(k)}=\bigtriangledown \times {\bf V}^{(k)}
\label{9}
\end{eqnarray}
are the $\omega-$ and $\rho-$ magnetic field strength respectively, and
\begin{eqnarray}
{\bf j}=\Psi^\dagger \vec \alpha \Psi \;\;\;\;\mbox{and}\;\;\;\;
{\bf j}^{(k)}=\Psi^\dagger \vec \alpha \tau^{(k)}\Psi  \label{10}
\end{eqnarray}
are the nucleon current density and nucleon isotopic
spin current density vectors.

Now let us quantize the nucleon field $\Psi$. At the zero
temperature, the proton number density and the neutron number density are
\begin{eqnarray}
n_P=\frac{p_P^3}{3\pi^2}\;\;\;\;\mbox{and}\;\;\;\;
n_N=\frac{p_N^3}{3\pi^2},\label{11}
\end{eqnarray}
respectively, with corresponding Fermi momenta $p_P$ and $p_N$ defined in it.
The nucleon number density and the third component of the nucleon
isotopic spin density are
\begin{eqnarray}
n=n_P+n_N \;\;\;\;\mbox{and}\;\;\;\;n^{(3)}=n_P-n_N=-In, \label{12}
\end{eqnarray}
in which $I=(n_P-n_N)/n$ is the isotopic asymmetry parameter.
We may also introduce the Fermi momentum p for the nucleon by
\begin{eqnarray}
n=\frac{2p^3}{3\pi^2}.     \label{i1}
\end{eqnarray}
There are simple relations
\begin{eqnarray}
p_P=(1-I)^{1/3}p,\;\;\;\;\mbox{and}\;\;\;\;p_N=(1+I)^{1/3}p. \label{i2}
\end{eqnarray}

In nuclear matter the effective nucleon mass becomes
\begin{eqnarray}
    m'=m-g\phi=m\chi,                    \label{13}
\end{eqnarray}
because of its interaction with the scalar meson field $\phi$.
\begin{eqnarray}
     \chi=\frac{m'}{m}=1-\frac{g\phi}{m}   \label{14}
\end{eqnarray}
is the effective mass of a nucleon in unit of the free nucleon mass $m$.
The average energy of a nucleon is
\begin{eqnarray}
\epsilon_1=\frac{1-I}{2}F(p_P,\chi)+\frac{1+I}{2}F(p_N,\chi)   \label{15}
\end{eqnarray}
with
\begin{eqnarray}
F(p,\chi)=\frac{3}{4} \left[ (1+\frac{\chi^2}{2p^2})\sqrt{p^2+\chi^2}
- \frac{\chi^4}{2p^3}\ln\left(\frac{p}{\chi}+\sqrt{1+\frac{p^2}{\chi^2}}\
\end{eqnarray}
Here we use the nucleon unit system, in which the unit of energy
is the static energy $mc^2$ of the nucleon, the unit of momentum is $mc$
, and the unit of length is
the nucleon Compton wave length $\lambda_c =\hbar /mc$.

The mass of the scalar meson
is defined in the coefficient $m_s^2/2$ of the $\phi^2$ term
in the potential energy density $U(\phi)$ for the scalar meson field.
{}From (\ref{14})
we may express $U$ in the form
\begin{eqnarray}
U=\frac{1}{3\pi^2\alpha}(1-\chi)^2[1+\alpha_1 (1-\chi)+\alpha_2(1-\chi)^2] ,
\label{17}
\end{eqnarray}
in which
\begin{eqnarray}
\alpha= \frac{2g^2m^2}{3\pi^2m_s^2}.   \label{18}
\end{eqnarray}
In the static homogeneous nuclear matter $\partial_\mu =0$,
and all 3-vectors are zero.
For given density $n$ and given charge state $I$ the third component
of the isotopic spin density is fixed by (\ref{12}).
But we may set $n^{(1)}=n^{(2)}=0$
to minimize the energy.
{}From (\ref{8}) we see the energy per-nucleon
in ground state nuclear matter is
\begin{eqnarray}
\epsilon=\epsilon_1+ \frac{U}{n}+2\pi\alpha_\omega
\frac{m^2}{m_\omega^2}n+2\pi\alpha_\rho \frac{m^2}{m_\rho^2}I^2n \label{19}
\end{eqnarray}
with
\begin{eqnarray}
\alpha_\omega =\frac{g_\omega^2}{4\pi}\;\;\;\;\mbox{and}\;\;\;\;
\alpha_\rho=\frac{g_\rho^2}{4\pi} \label{20}
\end{eqnarray}
being the $\omega-$ and $\rho-$ fine structure constants respectively.
$p_P$ and $p_N$ are related to $n$ and $I$ by (\ref{11}) and (\ref{12}).
Therefore, for given model parameters
\begin{eqnarray}
\alpha,\;\; \alpha_\omega,\;\; \alpha_\rho,\;\;
\alpha_1\;\; \mbox{and}\;\; \alpha_2 \label{i3}
\end{eqnarray}
the energy per-nucleon $\epsilon$ in nuclear matter
is a function of nucleon number density $n$,
isotopic spin asymmetry parameter $I$ and the
Minimizing the energy for given $n$ and $I$, we may also fix $\chi$.

\begin{center}
B. Extended nucleons interacting with the meson field
\end{center}

In Van der Waals approximation the extended structure of nucleons is
considered by simply cutting  out the volume occupied
by nucleons from the configuration space.
It changes the relation (\ref{11}) between the number density
and the Fermi momentum.
This relation is from the expression of the state number
\begin{eqnarray}
\Omega =2\frac{{\cal V} d^3 p}{(2\pi)^3} \label{21}
\end{eqnarray}
for one-half spin particles in the volume $\cal V$ and the momentum element
$d^3 p$. If the nucleon has a finite volume $v=4\pi a^3/3$, the
effective volume in which particles can move becomes
${\cal V}'={\cal V}-Nv, N$ is
the total nucleon number. Correspondingly, the state number becomes
\begin{eqnarray}
\Omega'=2\frac{{\cal V'} d^3 p}{(2\pi)^3}=(1-nv)\Omega,      \label{22}
\end{eqnarray}
and the relations between the number densities and the Fermi momenta
become
\begin{eqnarray}
n_P=(1-nv)\frac{p_P^3}{3\pi^2},\;\;\;\;
n_N=(1-nv)\frac{p_N^3}{3\pi^2}\;\;\;\;\mbox{and}\;\;\;\;
n=(1-nv)\frac{2p^3}{3\pi^2},     \label{23}
\end{eqnarray}
with (\ref{i2}) unchanged.
Solving the last equation we obtain
\begin{eqnarray}
n=\frac{2p^3}{3\pi^2+2v p^3},     \;\;\;\;\;\;\;\;\;\;
1-nv=\frac{3\pi^2}{3\pi^2+2v p^3}.  \label{24}
\end{eqnarray}
Since the same factor $(\!1\!-\!n\!v)$
also appears in the expression for energy density $\cal E$,
the energy per-nucleon
(\ref{19}), with (\ref{15}-\ref{16}) does not change.
It is still a function of $n,I,$ and $\chi$
with $\chi$ fixed by the energy minimization.
Only the relations between the densities and the Fermi momenta
are changed to (\ref{23}) with  (\ref{24}).

\begin{center}
C. Model parameters
\end{center}

Beside the five parameters in (\ref{i3}) we have one more parameter $a$,
the radius
of the extended baryon.
There is a constraint
\begin{eqnarray}
\alpha_1^2 \leq 4\alpha_2  \label{25}
\end{eqnarray}
to keep the vacuum being stable at  $\chi=1$.
We put $\alpha_1^2= 4\alpha_2$ for simplicity.
There are still five independent parameters.
They are going to be fixed by fitting
four empirical data\cite{ms}, namely the equilibrium
density $n_0=\left(\frac{4\pi}{3}r^3\right)^{-1}$ with r=1.175$fm$,
the binding energy per-nucleon $B/A=15.986MeV$,
the asymmetry energy $E_{as}=36.5MeV$ and the
compression modulus $K=240MeV$, for normal nuclear matter.
The reasonable range for the effective nucleon mass $\chi$
in normal nuclear matter
is $0.7-0.9$. It is also a constraint on the model parameters.

A fit of four data with five parameters is not unique.
It seems that we have some freedom to chose
one parameter. The constraint on $\chi$ limits this freedom.
We indeed have found a series of
parameter sets, every set of those six parameters fits
the above mentioned empirical nuclear matter data as well.
Among the sets we show the simplest one:
\begin{eqnarray}
\alpha_1=2.00,\;\alpha_2=1.00,\;\alpha=12.4,\;
\alpha_\omega=3.94,\;\alpha_\rho=1.83,\;a=0.62fm,  \label{26}
\end{eqnarray}
with $\chi=0.84$ in the normal nuclear matter.
We see they are quite reasonable.

\begin{center}
III. Equation of state for neutron star matter

\vskip 0.1in
\end{center}

Neutron star matter is a neutral matter composed of neutrons and other
hadrons and leptons in complete equilibrium at zero temperature. In this
work, we consider the lowest baryon octet, the electron, the muon,
the scalar meson field the $\omega-$ and $\rho-$ meson fields. Assuming
that the baryon masses are generated by the scalar meson field, we arrive to
the conclusion that the coupling constant $g_b$ of the scalar meson with
the baryon of the kind $b$ is proportional to the mass $m_b$ of that baryon.
The coupling of the $\omega-$ and $\rho-$ meson fields with various baryons
is assumed to be universal.

(\ref{23}) can be easily generalized to any Baryon. The number density of
baryon $b$ is
\begin{eqnarray}
n_b=(1-nv)\frac{p_b^3}{3\pi^2},\;\;
\mbox{with the total baryon number density}\;\;
n=\sum_{b=1}^8 n_b.   \label{27}
\end{eqnarray}
$p_b$ is the Fermi momentum for
baryon $b$. We may also introduce the Fermi momentum $p$
for the baryon octet by generalizing the last equation of (\ref{23}) to
\begin{eqnarray}
n=(1-nv)\frac{8p^3}{3\pi^2}  ,\label{i4}
\end{eqnarray}
We may solve from this equation
\begin{eqnarray}
n=\frac{8p^3}{3\pi^2+8vp^3} , \;\;\;\;\mbox{and}\;\;\;\;
1-nv=\frac{3\pi^2}{3\pi^2+8vp^3}.\label{i5}
\end{eqnarray}
Because of large mass differences for baryons in the octet, the Fermi
momentum for the octet is not meaningful. The first equation in (\ref{i5})
may be simply considered as a transformation for the density $n$. $p$ has
the advantage in that it may approach infinity, while $n\leq v^{-1}$.
The number density of the lepton $l$ is simply
\begin{eqnarray}
n_l=\frac{p_l^3}{3\pi^2},  \label{28}
\end{eqnarray}
$p_l$ is its Fermi momentum.
The energy density of the neutron star matter is
\begin{eqnarray}
{\cal E}=\sum_{b=1}^8n_bF(p_b,\frac{m_b}{m}\chi)
+\sum_{l=1}^2n_lF(p_l,\frac{m_l}{m})
      +U+2\pi(\alpha_\omega\frac{m^2}{m_\omega^2}n_\omega^2+\alpha_\rho
      \frac{m^2}{m_\rho^2}n_\rho^2). \label{29}
\end{eqnarray}
Again, the energy unit is $mc^2$, and the length unit is $\lambda_c$.
$n_\omega$ is the density generating the $\omega-$ field, and $n_\rho$
is the density generating the $\rho-$ field.
For the universal coupling to the vector mesons, we have
\begin{eqnarray}
n_\omega =n\;\;\;\;\mbox{and}\;\;\;\; n_\rho=\sum_{b=1}^8 A_b n_b, \label{30}
\end{eqnarray}
with
\begin{eqnarray}
A_b=\left\{ \begin{array}{ll} 2\;\;\;\;
& \mbox{for $b=\Sigma^+$}\\ 1\;\;\;\;
&\mbox{for $b=\Xi^0$ and $P$}\\ 0\;\;\;\;&
\mbox{for $b=\Lambda$ and $\Sigma^0$}\\ -1\;\;\;\;
&\mbox{for $b=\Xi^-$ and $N$}\\
-2\;\;\;\;&\mbox{for $b=\Sigma^-$}   \end{array} \right..\label{31}
\end{eqnarray}
We may solve the set of Fermi momenta $[p_b, p_l]$ from the set of densities
$[n_b, n_l]$, therefore the energy density $\cal E$ is a function of
densities $n_b, n_l$ and the effective mass $\chi$ of nucleon.
Since the later
may be fixed by the minimization condition
$\frac{\partial {\cal E}}{\partial \chi}=0$,
for $\cal E$, the energy density $\cal E$ becomes a
function of densities only.
{}From this function we obtain the pressure
\begin{eqnarray}
{\cal P}\!&\!=\!&\!\sum_{b=1}^8n_b\frac{\partial {\cal E}}{\partial n_b}
+\sum_{l=1}^2 n_l\frac{\partial {\cal E}}{\partial n_l} -
{\cal E} \nonumber \\
&\!=\!&\!\sum_{b=1}^8n_bH(p_b,\frac{m_b}{m}\chi)
+\sum_{l=1}^2n_lH(p_l,\frac{m_l}{m})
      -U+2\pi(\alpha_\omega\frac{m^2}{m_\omega^2}n_\omega^2+\alpha_\rho
      \frac{m^2}{m_\rho^2}n_\rho^2), \label{32}
\end{eqnarray}
with
\begin{eqnarray}
H(p,\chi)=\frac{1}{12\pi^2}\left[p(p^2-\frac{3}{2}\chi^2)\sqrt{p^2+\chi^2}
+\frac{3}{2} \chi^4\ln\left(\frac{p}{\chi}
+\sqrt{1+\frac{p^2}{\chi^2}}\right)\right]. \label{33}
\end{eqnarray}
The electric charge number density is
\begin{eqnarray}
n_c=\sum_{b=1}^8 C_b n_b-n_e-n_\mu,  \label{34}
\end{eqnarray}
with
\begin{eqnarray}
C_b=\left\{\begin{array}{ll} 1 \;\;\;\;& \mbox{for $b=P$ and $\Sigma^+$}
\\ 0\;\;\;\; & \mbox{for $b=N,\Lambda,\Sigma^0.$ and $\Xi^0$}
\\ -1\;\;\;\;
& \mbox{for $b=\Sigma^-$ and $\Xi^-$}   \end{array}  \right.. \label{35}
\end{eqnarray}

Now let us minimize the energy density under fixed baryon number
density $n$ and charge
number density $n_c$. It is to minimize the free energy
\begin{eqnarray}
  {\cal F}={\cal E}-\mu n +\mu_c n_c. \label{36}
\end{eqnarray}
$\mu$ and $\mu_c$ are Lagrange multipliers for conserved
baryon number and electric
charge respectively. From $\frac{\partial {\cal F}}{\partial p_b}=0$
we have
\begin{eqnarray}
\frac{p_b^2}{\pi^2}(\sqrt{p_b^2+\frac{m_b^2}{m^2}\chi^2}-\mu'+\mu_\rho A_b
+\mu_c C_b)=0.         \label{37}
\end{eqnarray}
$\mu'$ is the effective Lagrange multiplier for $n$, related to $\mu$.
However, it is directly determined by $n$, therefore we do not need  the
relation between $\mu$ and $\mu'$. $\mu_\rho$ looks like a Lagrange
multiplier for $n_\rho$. But it is directly related to $n_\rho$ by
\begin{eqnarray}
\mu_\rho =4\pi \alpha_\rho\frac{m^2}{m_\rho^2} n_\rho , \label{38}
\end{eqnarray}
therefore is not independent. Since the third component of isotopic
spin of baryons is not conserved in weak interaction, there is no
true Lagrange multiplier for $n_\rho$.
The solution of (\ref{37}) is
\begin{eqnarray}
p_b=\left\{ \begin{array}{ll}
\sqrt{(\mu'-\mu_\rho A_b-\mu_c C_b)^2-m_b^2 \chi^2 / m^2}
& \mbox{if $\mu'>\mu_\rho A_b +\mu_c C_b + m_b\chi / m $ } \\
0& \mbox{otherwise}. \end{array} \right. \label{39}
\end{eqnarray}
{}From $\frac{\partial {\cal F}}{\partial p_l}=0$ we have
\begin{eqnarray}
\frac{p_l^2}{\pi^2}\left(\sqrt{p_l^2+\frac{m_l^2}{m^2}}-\mu_c\right)
=0.\label{40}
\end{eqnarray}
Its solution is
\begin{eqnarray}
p_l=\left\{\begin{array}{ll} \sqrt{\mu_c^2 - m_l^2/m^2}
& \mbox{if $\mu_c > m_l/m$}\\
0 & \mbox{otherwise}.
\end{array}\right.   \label{41}
\end{eqnarray}
Substituting the obtained Fermi momenta $p_b$ and $p_l$
in (\ref{27}), (\ref{30}) and(\ref{34}),
we may calculate $n, n_\rho,$ and $n_c$ for given $n, \mu',
\mu_\rho$ and $\mu_c$. With $n_\rho$ we may also calculate
$\mu_\rho$ from (\ref{38
Putting $n_c=0$, requiring the calculated $n$ and $\mu_\rho$
be the same that we have put in the equation,
we may solve the Lagrange multipliers $\mu',
\mu_\rho$ and $\mu_c$ from a single
given number density $n$.
Therefore everything,
including $\cal E \mbox{and} P$ is determined by the total baryon number
density $n$. A relation between $\cal P$ and $\cal E$ is found.
That is the equation of state for the neutron star matter.
The numerical equation of state in the model with
the set (\ref{26}) of parameters is shown in Fig.\ 1.

\begin{center}
IV. Neutron Star Mass Limit
\vskip 0.1in
\end{center}

Now we are going to solve the Oppenheimer-Volkov equation\cite{w}
\begin{eqnarray}
\frac{d{\cal P}}{dr} &
=& -\frac{GM(r){\cal E}(r)}{r^2}
\left(1+\frac{{\cal P}(r)}{{\cal E}(r)}\right)
\left(1+\frac{4\pi r^3 {\cal P}(r)}{M(r)}
\right) \left(1-\frac{2GM(r)}{r}\right)^{-1}, \label{42} \\
\frac{dM}{dr} &=& 4\pi r^2{\cal E}(r),   \label{43}
\end{eqnarray}
in which $M(r)$ is the mass in the sphere of radius r. For a given
equation of state $\cal E(P)$, it is a set of well defined equations.
We have to solve it under the initial condition $M(0)=0$
and ${\cal P}(0)$ or
${\cal E}(0)$ being given, and to the radius where the pressure $\cal P$
may be ignored. In the following we would use the unit system in which
the unit of length is $Km$, the unit of mass is the solar mass $M_\odot$,
and the unit of energy is $M_\odot c^2$. In this unit system,
the gravitation constant
is $G=1.475$.
The transformation of $\cal E$ and $\cal P$ from the old unit system to
this system is simply multiplying a factor $9.197\times 10^{-2}$.

Using the equation of state obtained in the last section (shown in Fig.\ 1)
we solved the set of equations (\ref{42}) and (\ref{43}) numerically for
a set of central densities ${\cal E}(0)$. The results are shown in Fig.\ 2.
We see that the maximum mass is over $2.5M_\odot$ It is appreciably larger
than the mass $1.85M_\odot$ of the observed heaviest neutron star 4U0900-40.
This result confirms our idea expressed in the introduction.
That is after the finite baryon volume being considered,
we may find a equation of state for hadron matter,
which is consistent with both
the normal nuclear data including the compression modulus $K=240MeV$
and the data for
high density hadron matter such as the mass limit of neutron star.
\newpage
\begin{center}
V. Discussion
\vskip 0.1in
\end{center}

Our idea is simple: The finite volume of baryons may make the equation
of state for hadron matter stiffer at high density whereas it is kept
consistent with all normal nuclear data
including the low compression modulus $K=240MeV$ at normal nuclear density.
The stiff equation of state seems needed for understanding various phenomena
at high density both in heavy ion collision and in neutron star.
We confirmed
this idea by numerical calculation of the mass limit for neutron star here.
The success encourages us to investigate similar effects
in the heavy ion collision phenomena.
Since we are only interested in the role of finite volume effect of baryons,
some simplification has been made. We assumed that the volume of various
baryons are the same. In reality, the volumes of hyperons should be bigger
than that of nucleon. If we consider this point, the effect should be even
larger. We also ignored an important baryon, that is the $\Delta$-resonance.
{}From the physics consideration behind the calculation,
we believe this point
could not change the qualitative result.

Of course, there is something over simplified in the present calculation.
The Van der Waals approximation
treats the finite volume of the baryon as a hardcore. It is essentially a
nonrelativistic idea. At very high density it may violate the causality.
Therefore we should not be too serious in considering
the quantitative result here.
We may put some criteria for the reliability of the calculated results.
If we put the so called causal limit $\cal P=E$ as the criterion,
the result below
the density $n_{limit}=8n_0$ may be reliable. Under this limitation,
the mass limit for neutron stars is $2M_\odot$.
This is still larger than the observed neutron star masses.
The qualitative result is unchanged.
However, this discussion suggests a topic of further research: How to
handle the finite baryon volume effect for hadron matter
in a relativistic way.

The first author (QRZ) would acknowledge
Dr. J. Reinhardt for his help in computation.

\newpage
\begin{center}
Figure Captions
\end{center}
\vskip 0.5in
Fig.\ 1. Relation between the
pressure $\cal P$ and the energy density $\cal E$
in unit of $mc^2/\lambda_c^3$ for the neutron star matter
at zero temperature.
\vskip 0.2in
\noindent
Fig.\ 2. Relation between the mass $M$ in unit of
solar mass $M_\odot$ and the
central mass density $\cal N$ in unit of
the normal nuclear density $n_0m$ for the neutron stars.
\end{document}